\documentstyle[prl,aps,epsfig,multicol]{revtex}
\def\sea{$(\overline{d}(x)-\overline{u}(x))/(u(x)-d(x))\;\;$}
\def\seax{$(\overline{d}-\overline{u})/(u-d)\;\;$}
\def\seaxp{$(\overline{d}-\overline{u})/(u-d)$}

\def\seaox{$\overline{d}-\overline{u}\;\;$}
\def\seaoxp{$\overline{d}-\overline{u}$}

\def\seaoi{$\int_{0.02}^{0.3}(\overline{d}(x)-\overline{u}(x))dx\:$}

\def\seaoiq{$=0.107\pm0.021\rm{(stat)}\pm0.017\rm{(sys)}\,$}
\begin{document}
\draft
\title{The Flavor Asymmetry of the Light Quark Sea \\
from Semi-inclusive
  Deep-inelastic Scattering}
\author{
K.~Ackerstaff$^{5}$, 
A.~Airapetian$^{32}$, 
N.~Akopov$^{32}$,
I.~Akushevich$^{6}$ 
M.~Amarian$^{32,27}$, 
E.C.~Aschenauer$^{13,14}$, 
H.~Avakian$^{10}$, 
R.~Avakian$^{32}$, 
A.~Avetissian$^{32}$, 
B.~Bains$^{15}$, 
C.~Baumgarten$^{23}$,
M.~Beckmann$^{12}$, 
St.~Belostotski$^{26}$, 
J.E.~Belz$^{28,29}$,
Th.~Benisch$^8$, 
S.~Bernreuther$^8$, 
N.~Bianchi$^{10}$, 
J.~Blouw$^{25}$, 
H.~B\"ottcher$^6$, 
A.~Borissov$^{6,14}$, 
J.~Brack$^4$, 
S.~Brauksiepe$^{12}$,
B.~Braun$^{8}$, 
B.~Bray$^3$, 
St.~Brons$^6$,
W.~Br\"uckner$^{14}$, 
A.~Br\"ull$^{14}$, 
E.E.W.~Bruins$^{20}$,
H.J.~Bulten$^{18,25,31}$, 
R.V.~Cadman$^{15}$,
G.P.~Capitani$^{10}$, 
P.~Carter$^3$,
P.~Chumney$^{24}$,
E.~Cisbani$^{27}$, 
G.R.~Court$^{17}$, 
P.~F.~Dalpiaz$^9$, 
E.~De Sanctis$^{10}$, 
D.~De Schepper$^{20}$, 
E.~Devitsin$^{22}$, 
P.K.A.~de Witt Huberts$^{25}$, 
P.~Di Nezza$^{10}$,
M.~D\"uren$^8$, 
A.~Dvoredsky$^3$, 
G.~Elbakian$^{32}$, 
A.~Fantoni$^{10}$, 
A.~Fechtchenko$^7$,
M.~Ferstl$^8$, 
K.~Fiedler$^8$, 
B.W.~Filippone$^3$, 
H.~Fischer$^{12}$, 
B.~Fox$^4$,
J.~Franz$^{12}$, 
S.~Frullani$^{27}$, 
M.-A.~Funk$^5$, 
N.D.~Gagunashvili$^7$, 
H.~Gao$^{2,15}$,
Y.~G\"arber$^6$, 
F.~Garibaldi$^{27}$, 
G.~Gavrilov$^{26}$, 
P.~Geiger$^{14}$, 
V.~Gharibyan$^{32}$,
V.~Giordjian$^{10}$, 
A.~Golendukhin$^{19,32}$, 
G.~Graw$^{23}$, 
O.~Grebeniouk$^{26}$, 
P.W.~Green$^{1,29}$, 
L.G.~Greeniaus$^{1,29}$, 
C.~Grosshauser$^8$, 
A.~Gute$^8$, 
W.~Haeberli$^{18}$, 
J.-O.~Hansen$^2$,
D.~Hasch$^6$, 
O.~H\"ausser\cite{author_note1}$^{28,29}$, 
R.~Henderson$^{29}$, 
Th.~Henkes$^{25}$,
M.~Henoch$^{8}$, 
R.~Hertenberger$^{23}$, 
Y.~Holler$^5$, 
R.J.~Holt$^{15}$, 
W.~Hoprich$^{14}$,
H.~Ihssen$^{5,25}$, 
M.~Iodice$^{27}$, 
A.~Izotov$^{26}$
% ...
\centerline {\it (The HERMES Collaboration)}
}
\address{
% ...
$^5$DESY, Deutsches Elektronen Synchrotron, 22603 Hamburg, Germany\\
% ...
$^8$Physikalisches Institut, Universit\"at Erlangen-N\"urnberg, 
91058 Erlangen, Germany\\
% ...
$^{14}$Max-Planck-Institut f\"ur Kernphysik, 69029 Heidelberg, Germany\\ 
$^{15}$Department of Physics, University of Illinois, Urbana, 
Illinois 61801, USA\\
% ...
$^{23}$Sektion Physik, Universit\"at M\"unchen, 85748 Garching, Germany\\
% ...
$^{25}$Nationaal Instituut voor Kernfysica en Hoge-Energiefysica (NIKHEF), 
1009 DB Amsterdam, The Netherlands\\
$^{26}$Petersburg Nuclear Physics Institute, St. Petersburg, 188350 Russia\\
$^{27}$Istituto Nazionale di Fisica Nucleare, Sezione Sanit\'a and Istituto
Superiore di Sanit\'a, Physics Laboratory, 00161 Roma, Italy\\
$^{28}$Department of Physics, Simon Fraser University, Burnaby, 
British Columbia V5A 1S6, Canada\\ 
$^{29}$TRIUMF, Vancouver, British Columbia V6T 2A3, Canada\\
% ...
}
\date{\today}
\maketitle
\begin{abstract}
  
The flavor asymmetry of the light quark sea of the nucleon is determined 
in the kinematic range $0.02<x<0.3$ and 
$1$~GeV$^2<Q^2<10$~GeV$^2$,
for the first time
from semi-inclusive deep-inelastic scattering.
The quantity \sea is derived from a relationship between 
the yields of positive and negative
pions from unpolarized hydrogen and deuterium targets. 
The flavor asymmetry \seaox is found to be
non-zero and x dependent, showing an excess of $\overline{d}$ over 
$\overline{u}$ quarks
in the proton.
\end{abstract}
\pacs{13.87.Fh; 14.20.Dh; 14.65.Bt; 24.85.+p}

%HERE IS THE REVTEX COMMAND TO GET GALLEY FORMAT:
\begin{multicols}{2}[]
%\narrowtext
%\twocolumn

% body of paper here
The flavor content of the nucleon sea has come to be recognized as
an important domain for testing models of nucleon structure\cite{models}.
Until recently there has been little experimental constraint
on the flavor asymmetry in the
quark distributions of the light sea quarks in the nucleon.
The first was an integral test, 
based on a comparison of inclusive deep-inelastic
scattering on the proton and the neutron to determine the Gottfried
Sum\cite{Gott}
defined as
\begin{equation}
   S_G  \equiv \int_0^1\frac{dx}{x}\left(F_2^p(x)-F_2^n(x)\right),\nonumber
\end{equation}
where $F_2^p(x)$ and $F_2^n(x)$ are the structure functions of the proton
and neutron, respectively.
The assumption of isospin symmetry in the quark-parton model allows the 
Sum to be written as 
\begin{equation}
  S_G  =\frac{1}{3}\int_0^1(u_v(x)-d_v(x))dx
   -\frac{2}{3}\int_0^1(\overline{d}(x) -\overline{u}(x))dx,
\end{equation}
where $u_v(x)$ and $d_v(x)$ are the density
functions of the valence quarks and $\overline{u}(x)$ and $\overline{d}(x)$
those of the antiquarks in the proton,
and $x$ is the fraction of the nucleon light cone momentum
carried by the struck quark. 
A flavor symmetric sea, $\overline{u}(x)=\overline{d}(x)$,
leads to the Gottfried
Sum Rule $S_G=1/3$. A measurement of the Gottfried Sum 
by NMC~\cite{NMCGott} resulted in $S_G=0.235\pm 0.026$, at $Q^2$~=~4~GeV$^2$.
If isospin symmetry holds, a global flavor
asymmetry $\int_0^1(\overline{d}(x)-\overline{u}(x))dx\approx 0.15$
would account for the NMC result. Many ideas such as 
Pauli blocking and pion clouds have been proposed to explain such a flavor
asymmetry in the sea\cite{models}. 
Two methods have been proposed to measure its
$x$ dependence: the Drell-Yan process~\cite{EandS}
and semi-inclusive deep inelastic scattering~\cite{Levelt}. 

  Results are presented here
for the $x$ dependence of \seax from the analysis of
charged pion yields in
semi-inclusive deep-inelastic scattering of positrons on unpolarized
hydrogen and deuterium targets. 
The data cover the kinematic
range $0.02<x<0.3$ and 1~GeV$^2<Q^2<10$~GeV$^2$, where $-Q^2$ is
the four-momentum transfer squared of the exchanged virtual photon and
$x=Q^2/(2M\nu)$ is the Bjorken variable with $M$ being the proton mass,
$\nu=E-E'$ the virtual photon energy, and
$E(E')$ the energy of the incident (scattered) positron.

  The ratio \seax is determined from a quantity that combines good
sensitivity to the sea asymmetry with minimal 
sensitivity to instrumental effects.
This is the ratio of the differences between
charged pion yields for proton and neutron targets~\cite{Levelt}:
\begin{equation}
r(x,z)= \frac{N^{\pi^-}_p(x,z)-N^{\pi^-}_n(x,z)}{N^{\pi^+}_p(x,z)-
N^{\pi^+}_n(x,z)}, \label{eq:littler}
\end{equation}
where  $z~=~E^{\pi}/\nu$ is the fraction of the virtual photon energy carried
by the pion and $N^{\pi}(x,z)$ is the yield of pions coming from
deep-inelastic scattering off nucleons. Factorization between
the hard scattering process and the hadronization of the struck quark implies
\begin{equation}
N^{\pi^{\pm}}(x,z)\propto\sum_i e^2_i \left[q_i(x)D_{q_i}^{\pi^{\pm}}(z)+\overline{q}_i(x)D_{\overline{q}_i}^{\pi^{\pm}}(z)\right], \label{eq:mult}
\end{equation}
where  $e_i$ is the quark charge in units of the elementary charge,
$q_i(x)$ and $\overline{q}_i(x)$ are the density distributions 
of quarks and
antiquarks of flavor $i$, and the fragmentation
functions $D_{q_i}^{\pi^{\pm}}(z)$ represent the probability that the quark
of flavor $i$ fragments to a charged pion.
Assuming isospin symmetry between protons and neutrons as well as charge
conjugation invariance, the number of light quark fragmentation functions is
reduced to two: the favored and disfavored
fragmentation functions~\cite{EMCff}, e.g. $D_u^{\pi^+}$ and $D_u^{\pi^-}$. 
Using Eq.~\ref{eq:mult} to express $r(x,z)$ in terms of quark distributions 
and fragmentation functions results in
\begin{equation}
   \frac{1+r(x,z)}{1-r(x,z)}=\frac{u(x)-d(x)+\overline{u}(x)-\overline{d}(x)}
   {(u(x)-\overline{u}(x))-(d(x)-\overline{d}(x))}J(z), \label{eq:Rxz}
\end{equation}
where $J(z)=\frac{3}{5} \left(\frac{1+D'(z)}{1-D'(z)}\right)$ and
$D'(z)=D_u^{\pi^-}(z)/D_u^{\pi^+}(z)$.
It should be noted that the right-hand side of 
Eq.~\ref{eq:Rxz} factorizes into two independent functions of $x$ and $z$
respectively. Thus, the equation may be rearranged to isolate a quantity
sensitive to the flavor asymmetry:
\begin{equation}
   \frac{\overline{d}(x)-\overline{u}(x)}{u(x)-d(x)} =
    \frac{J(z)[1-r(x,z)]-[1+r(x,z)]}
         {J(z)[1-r(x,z)]+[1+r(x,z)]}.
   \label{eq:tautau}
\end{equation}
The assumptions leading to Eq.~\ref{eq:Rxz} imply that the expression
on the right side of Eq.~\ref{eq:tautau} is independent of $z$. 

  In the HERMES experiment, 27.5~GeV positrons circulating in the HERA storage
ring at DESY are scattered on internal hydrogen (H$_2$), deuterium (D$_2$), and
helium-3 ($^3$He) targets. The target gas is fed into the center of
a 40-cm-long open-ended storage cell. Systematic uncertainties are minimized 
by cycling the three gases through one of various permutations,
several times during each 8-12 hour HERA fill.
The unpolarized luminosity of $(1-5)\times 10^{32}$~cm$^{-2}$s$^{-1}$ 
is monitored by a pair of NaBi(WO$_4$)$_2$ electromagnetic
calorimeters that detect Bhabha scattering from target electrons. 

  A detailed description of the HERMES spectrometer is provided 
elsewhere~\cite{specpaper}. It is a forward spectrometer that 
identifies
the scattered positron as well as the associated hadrons, in the scattering 
angle  range of $0.04~\mbox{rad}<\theta<0.22$~rad.
Positron-hadron discrimination is based on information
from four particle identification detectors: a threshold gas 
\v{C}erenkov counter, a transition-radiation detector, a lead-glass
electromagnetic 
calorimeter, and a preshower detector located directly before the calorimeter.
This combination provides an average positron
identification 
efficiency of 99\%, with a hadron contamination that is dependent on the 
kinematics of the positron but is always less than 1\%. 
One of the features of the spectrometer central to this work is its threshold
\v{C}erenkov counter, which distinguishes pions from heavier hadrons 
for particles with momenta greater 
than 3.8~GeV while maintaining better than 95\% pion identification 
efficiency at momenta larger than 6~GeV.

The kinematic requirements used in this analysis are 
$Q^2>1$~GeV$^2$, invariant mass $W$ of the initial photon-nucleon 
system greater 
than 2~GeV, and fractional energy $y=\nu/E$ of the incident lepton 
transferred to the virtual photon less than 0.85.
Events are required to originate from the target-beam interaction region.
For the determination of $r(x,z)$, the data are
partitioned in five bins of $x$ in the
range $0.02<x<0.3$ and up to six bins in $z$ in the range $0.2<z<0.8$.
The pions are selected to be in
the current fragmentation region (i.e. derived from the struck quark
and not target fragments) by requiring $x_F=2P_L/W$ to be
greater than 0.25, where $P_L$ is the momentum component of the pion in the 
longitudinal direction with respect to the virtual
photon in the photon-nucleon center-of-mass frame. Data at Bjorken
$x>0.3$ are excluded because they are dominated by scattering from the
valence quarks. The yields for the neutron target appearing in
Eq.~\ref{eq:littler}
can be inferred from yields obtained from deuterium and hydrogen 
targets, $N^{\pi^{\pm}}_{n} =N^{\pi^{\pm}}_{d}-N^{\pi^{\pm}}_{p}.$
Thus the determination 
of \sea is made by taking ratios of differences between yields from 
hydrogen and deuterium. The pion
yields extracted from the data are ${N^{\pi}_{det}(x,z)}/{{\cal L}}$
where $N^{\pi}_{det}(x,z)$ is the number 
of detected pions associated with deep-inelastic scattering 
events, and ${\cal L}$ is the integrated
luminosity (nuclei/cm$^2$/s) 
corrected for instrumental dead-time and inefficiencies; the possible
effect of different acceptances for $\pi^+$ and $\pi^-$ is included
in the systematic uncertainties discussed below.
The fragmentation functions $D_{q_i}^{\pi^{\pm}}(z)$ 
and their ratio $D'(z)$ were extracted from the measured pion yields
from each of the unpolarised targets: H$_2$, D$_2$ and 
$^3$He~\cite{hermesff}, using Eq.~4 together with the 
GRV~94~LO~{\protect\cite{GRV}} parameterization of parton distributions.
No statistically significant differences were found between the results from
the different targets. To avoid statistical
correlations  
with the pion ratios, 
only $D'(z)$ extracted from
$^3$He is used in this analysis. 
The fragmentation function ratio $D'(z)$ determined within
the HERMES acceptance is found to vary
with $x$ and $Q^2$ typically by less than 10\%, except in the low-yield region
near $z=0.8$ where as much as 40\% may be seen.
However, extraction of the
flavor asymmetry in bins of $x$ and $Q^2$ revealed the averaged result
to be unaffected by this variation. The dependence of the left-right 
instrumental asymmetry on pion charge is significant only at the lowest
momenta and is taken into account as a correction. For all $x$ bins, the
resulting change in the sea asymmetry is less than 1.5\%.
 
\begin{figure}
   \epsfig{file=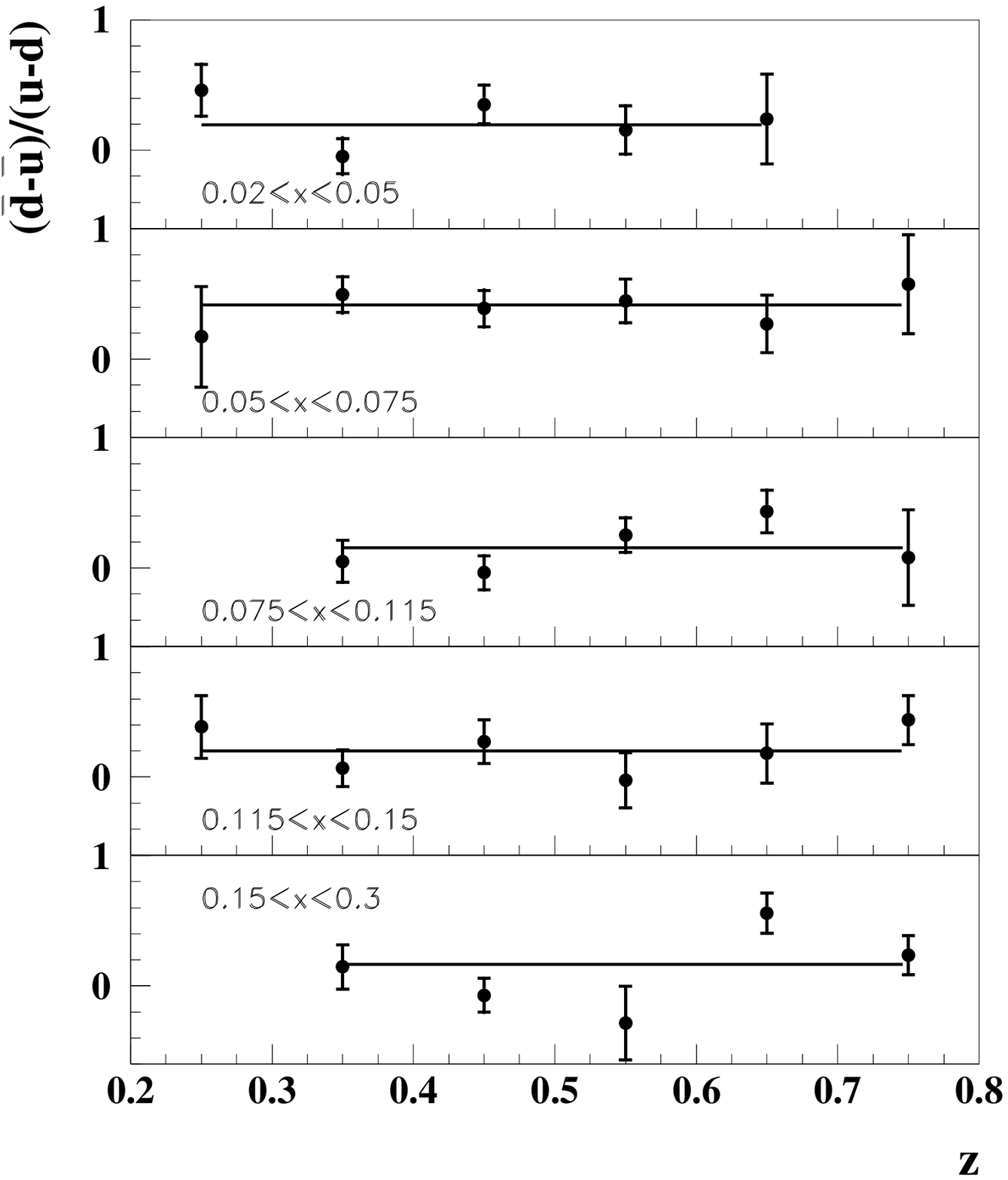,width=17cm}
      \caption{
The distribution $(\overline{d}-\overline{u})/(u-d)$ as a function of
  $z$ in five bins of $x$. The points are fit to a constant for each $x$ bin.
  The error bars represent statistical and systematic uncertainties
  added in quadrature. Some points are omitted due to limited statistics.
}
\label{fig:sea_z}
\end{figure}

\begin{figure}
   \epsfig{file=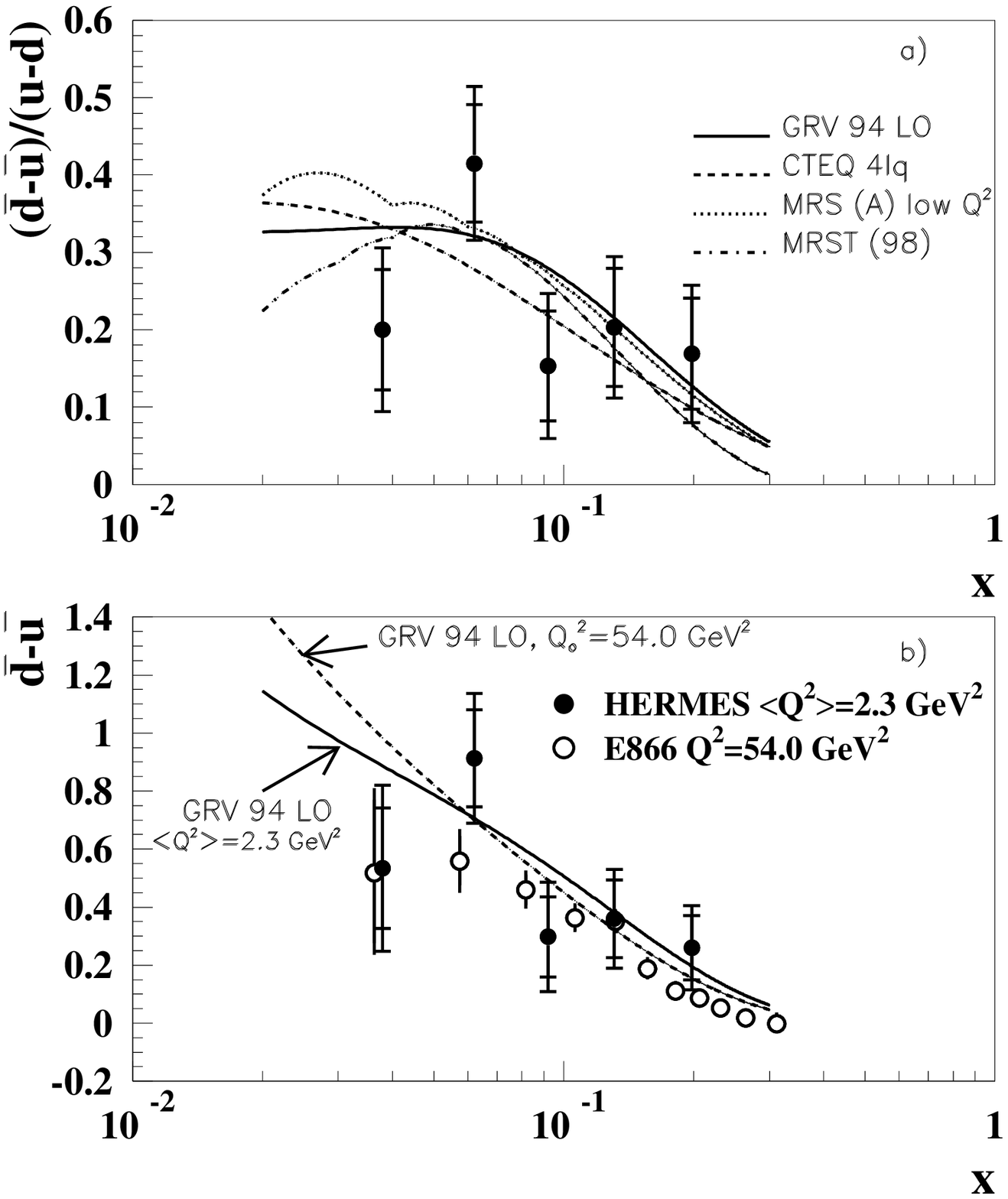,width=17cm}
      \caption{ (a) \seax as a function of $x$. Also included
   are the GRV~94~LO~{\protect\cite{GRV}},
   CTEQ~4lq~{\protect\cite{CTEQ4}}, MRS~(A)~{\protect\cite{MRSA}} low $Q^2$,
   and MRST (98)~{\protect\cite{MRS98}} parameterizations calculated at the 
   appropriate $Q^2$ for each $x$-bin. (b) \seaox as a function of $x$. 
   The open circles represent the E866~{\protect\cite{E866}} 
   determination of \seaoxp\ at $Q_0^2$=54~GeV$^2$. 
   The solid (dashed) curve is the 
   GRV~94~LO~{\protect\cite{GRV}} parameterization, evaluated at the $Q_0^2$
   values for the HERMES (E866) experiment.
The inner error bars represent the statistical uncertainties while
for all data including those from E866, the total 
error bars represent statistical and systematic
uncertainties added in quadrature.
}
\label{fig:sea_x}
\end{figure}

The results for the quantity \seax are shown in Fig.~\ref{fig:sea_z}
as a function of $z$ for five bins in $x$. The data show no
$z$ dependence and are consistent with the form of factorization
shown in Eq.~\ref{eq:Rxz}. The values of \seax averaged over $z$
are plotted in
Fig.~\ref{fig:sea_x}(a) as a function of $x$,  and are presented in 
Table~\ref{tab:errorsea}. They are non-zero and positive everywhere
in the measured $x$ region,  clearly showing
an excess of $\overline{d}$ quarks over
$\overline{u}$ quarks in the proton. Also included in this figure are the 
GRV~94~LO~{\protect\cite{GRV}}, CTEQ~4lq~{\protect\cite{CTEQ4}},
MRS~(A)~\cite{MRSA} low $Q^2$, and MRST (98)~\cite{MRS98}
parameterizations of \seaxp, calculated at the average $Q^2$ of the events in
each $x$ bin. 
The HERMES results are determined without making higher order
QCD corrections and so only leading order parameterizations should be
compared to them.
At low $Q^2$, 
only GRV~94 is available at leading order.
The distribution \seaox as a function of $x$ is thus derived from \seax
using the GRV~94~LO  parameterization 
of $u(x)-d(x)$, again calculated at the appropriate $Q^2$ for each $x$ bin.
Other combinations from GRV~94~LO could have been used instead, such as
$u_v-d_v$ or $u+\overline{u}-d-\overline{d}$, but these choices result
in more sensitivity to this model input, as indicated by differences
seen from substituting the GRV~98~LO parameterization\cite{GRV98}.
The results are presented in Fig.~\ref{fig:sea_x}(b) and
Table~\ref{tab:errorsea}, and give an integral 
over the measured 
$x$ region of \seaoi\seaoiq,
after the integrand at each x bin is `evolved' to the average measured 
$Q_o^2$ of 2.3~GeV$^2$ by applying the ratio of the GRV~94 values for 
\seaox at the two values of $Q^2$. 
The GRV~94~LO parameterization is used to estimate the
portion of the integral in the unmeasured regions. 
The contribution from $x<0.02$
is significant: 0.050 
(almost 50\% of that over the measured $x$ range), while the high $x$ region $x>0.3$ 
contributes only 0.006.
Including an extrapolation uncertainty estimate based on
differences among the available parameterizations, the total
integral over all $x$ is then approximately $0.16\pm 0.03$, which is
consistent with the results first seen by NMC\cite{NMCGott}.

An alternate approach to the measurement of the flavor asymmetry is the
comparison of
the Drell-Yan process on protons and deuterons, which is sensitive to the
quantity $\overline{u}/\overline{d}$~\cite{EandS}. Using this approach,
the NA51 collaboration at CERN determined 
that $\overline{u}/\overline{d}=0.51\pm 0.04\pm 0.05$
at $x=0.18$ and at a center-of-mass energy of $\sqrt{s}=29$~GeV~\cite{NA51}. 
The E866 collaboration at 
Fermilab recently reported a measurement of the flavor asymmetry in the $x$ 
range $0.02<x<0.345$~\cite{E866}.
E866 uses the CTEQ~4M~{\protect\cite{CTEQ4}}
parameterization of $\overline{u}+\overline{d}$ to extract \seaox from
the measured quantity
$\overline{d}/\overline{u}$. These results given at a $Q_o^2$ of 54~GeV$^2$ 
are included in  Fig.~\ref{fig:sea_x}(b).
This figure demonstrates that the sea asymmetry measured in deep-inelastic
scattering and in Drell-Yan experiments agree,
even though the $Q^2$ of the two experiments differ by a factor of
about 20.

The total systematic uncertainties associated with the determination of the
quantity \seax in the present work are given in Table~\ref{tab:errorsea}.
These uncertainties include various effects as follows. 
After the correction at low momenta,
any residual charge-dependent instrumental acceptance asymmetry
is estimated to be less than 1.5\%. Effects of the finite resolution of the
spectrometer and bremsstrahlung have been investigated with Monte Carlo 
studies. These effects are small and are included in the systematic
uncertainty. Radiative corrections to the pion yields 
have been calculated following the presciption of reference~\cite{radcor}.
Their effect on the pion ratio in Eq.~\ref{eq:littler} is estimated to be less
than 1\% and is included in the systematic uncertainty.
Lepton contamination in the pion sample is found to be at most of order 
one percent, and the
differences between the two targets were no more than 0.2\%. The effects 
of hadrons misidentified as pions have been estimated from the Monte Carlo
simulations and are also included. Finally, the uncertainty in the 
fragmentation function ratio $D'(z)$ as measured at HERMES has been
accounted for. The statistical uncertainty of $D'(z)$ has been included in
quadrature to the statistical uncertainty of the pion yields. The systematic
component of this ratio determined from systematic studies of the HERMES
data has been included in quadrature to the total uncertainty.

The only additional systematic uncertainty associated with \seaox is the
choice of parameterization to extract \seaox from \seaxp. 
The GRV~94~LO~{\protect\cite{GRV}} parameterization of $u(x)-d(x)$ is used.
The MRS~(A)~\cite{MRSA} low
$Q^2$ parameterization is used as an alternative.
The resulting difference in the value of \seaox is
used as an estimate of this uncertainty.

  In summary, the flavor asymmetry of the light quark sea is extracted
for the first time from pion yields in semi-inclusive deep-inelastic 
scattering. The data are recorded at the HERMES experiment in the 
kinematic range of $0.02<x<0.3$ and 
1~GeV$^2<Q^2<10$~GeV$^2$. 
The measured quantity \seax is found to be non-zero over
the entire $x$ range measured, clearly showing an excess of
$\overline{d}$ quarks over $\overline{u}$ quarks in the
proton sea. Its $x$ dependence will serve as a further
constraint on global parameterizations of parton density
functions. The integral over the measured region of the
derived quantity \seaox is \seaoi\seaoiq\ at a $Q_o^2$ 
of 2.3~GeV$^2$, which accounts for two-thirds
of the Gottfried Sum Rule deficit.

We gratefully acknowledge the DESY management for its support and 
the DESY staff and the staffs of the collaborating institutions. 
This work was supported by  
the FWO-Flanders, Belgium;
the Natural Sciences and Engineering Research Council of Canada;
the INTAS, HCM,  and TMR network contributions from the European Community;
the German Bundesministerium f\"ur Bildung, Wissenschaft, Forschung
und Technologie; the Deutscher Akademischer Austauschdienst (DAAD);
the Italian Istituto Nazionale di Fisica Nucleare (INFN);
Monbusho, JSPS, and Toray
Science Foundation of Japan;
the Dutch Foundation for Fundamenteel Onderzoek der Materie (FOM);
the U.K. Particle Physics and Astronomy Research Council; and
the U.S. Department of Energy and National Science Foundation.

% now the references. delete or change fake bibitem. delete next three
%   lines and directly read in your .bbl file if you use bibtex.

\end{multicols}

% \begin{table}
% \caption{}
% \label{}
% \end{tabular}
% \end{table}

\begin{table}
\caption{ Values for \seax and \seaoxp.}
\label{tab:errorsea}
\begin{tabular}{|c|c|c||c|c|c||c|c|c|} \hline
$x$         & $<x>$ & $<Q^2>$ & $\underline{\overline{d}-\overline{u}}$ & Stat. & Sys. & $\overline{d}-\overline{u}$  & Stat.  & Sys.\\
range       &       &(GeV$^2$)&  $u-d$                                  & error& error &                              & error  & error    \\ \hline
0.020-0.050 & 0.038 & 1.33 & 0.20 & 0.08 & 0.07 & 0.53 & $ 0.21$ & $ 0.20$   \\
0.050-0.075 & 0.062 & 1.82 & 0.42 & 0.08 & 0.06 & 0.91 & $ 0.17$ & $ 0.15$ \\
0.075-0.115 & 0.092 & 2.52 & 0.15 & 0.07 & 0.06 & 0.30 & $ 0.14$ & $ 0.13$ \\
0.115-0.150 & 0.131 & 3.38 & 0.20 & 0.08 & 0.05 & 0.36 & $ 0.13$ & $ 0.10$ \\
0.150-0.300 & 0.198 & 4.88 & 0.17 & 0.07 & 0.05 & 0.26 & $ 0.11$ & $ 0.09$  \\ \hline
\end{tabular}
\end{table}
% \begin{figure}
% \caption{}
% \label{}
% \end{figure}

% ****** End of file template.aps ******

%* *********************************************************************


\begin{references}
\bibitem[*]{author_note1} Deceased.
\bibitem{models}        For a recent review see F.M. Steffens and 
                        A. W. Thomas, Phys. Rev.
                        {\bf C55}, 900 (1997) and references therein.
\bibitem{Gott}          K. Gottfried, Phys. Rev. Lett. {\bf 18}, 1174 (1967). 
\bibitem{NMCGott}       NMC Collaboration,  M. Arneodo {\it et al}.,
                        Phys. Rev. {\bf D50}, 1 (1994).
\bibitem{EandS}         S. D. Ellis and W. J. Stirling,
                        Phys. Lett. {\bf B256}, 258 (1991).
\bibitem{Levelt}        J. Levelt {\it et al}., Phys. Lett. {\bf B263}, 498
                        (1991).
\bibitem{EMCff}         EMC Collaboration, M. Arneodo {\it et al}.,
                        Nucl. Phys. {\bf B321}, 541 (1989).
\bibitem{specpaper}     HERMES Collaboration, K. Ackerstaff {\it et al}.,
                        DESY Report 98-057, 
                        http://www.desy.de/library/preprints.html,
                        accepted by Nucl. Instr. and Meth.
\bibitem{hermesff}      P. Geiger, Ph.D. thesis,
                        Ruprecht-Karls-Universit\"{a}t, Heidelberg 1998,
                http://dxhra1.desy.de/notes/pub/98-LIB/geiger.98.005.ps.gz. 
\bibitem{GRV}           M. Gl\"{u}ck {\it et al}.,
                        Z. Phys. {\bf C67}, 433 (1995).
\bibitem{GRV98}         M. Gl\"{u}ck {\it et al}.,
                        DO-TH-98-07, hep-ph/9806404 (1998).
\bibitem{CTEQ4}         H. L. Lai {\it et al}.,
                        Phys. Rev. {\bf D55}, 1280 (1997).
\bibitem{MRSA}          A. D. Martin {\it et al}.,
                        Phys. Rev. {\bf D51}, 4756 (1995).
\bibitem{MRS98}         A. D. Martin {\it et al}.,
                        hep-ph/9803445.
\bibitem{NA51}          NA51 Collaboration, A. Baldit {\it et al}.,
                        Phys. Lett. {\bf B332}, 244 (1994).
\bibitem{E866}          E866 Collaboration, E. A. Hawker {\it et al}.,
                        Phys. Rev. Lett. {\bf 80}, 3715 (1998);
                        E866 Collaboration, J.C. Peng {\it et al}.,
                        hep-ph/9804288.
\bibitem{radcor}        T. V. Kukhto and N. M. Shumeiko, Nucl. Phys. 
                        {\bf B219},
                        412 (1983); I. V. Akushevich and N. M. Shumeiko, 
                        J. Phys. {\bf G20}, 513 (1994);
                        A. V. Soroko and N. M. Shumeiko in {\sl Workshop on 
                        Radiative Corrections Relevant for the HERMES 
                        Experiment}, edited by H. Boettcher and W.-D. Nowak
                        [Report No. DESY-Zeuthen 94-02].

\end{references}
\end{document}